\documentclass[prm,10pt,floatfix,amsmath,twocolumn,showpacs,superscriptaddress]{revtex4-1}
\usepackage{latexsym}
\usepackage{amsmath}
\usepackage{amssymb}
\usepackage[ansinew]{inputenc}
\usepackage{graphicx}
\usepackage{pst-all}
\usepackage{color}
\usepackage{dcolumn}
\usepackage{amsthm}
\usepackage{amsmath}
\usepackage{bm}
\usepackage{diagbox}
\usepackage{layout}
\usepackage{booktabs}
\usepackage{amsfonts}

\makeatletter

\newcommand{\Rmnum}[1]{\expandafter\@slowromancap\romannumeral #1@}
\makeatother

\usepackage[colorlinks=true,linkcolor=blue,
pagecolor=blue,filecolor=blue,
menucolor=blue,urlcolor=blue,
citecolor=blue,anchorcolor=blue]{hyperref}%

\begin{document}

\title{
Electronic structures and stability investigation of the new class of large band gap topological insulators $M$Tl$_4$Te$_3$ ($M$ = Cd, Hg)
}

\author{ Ying Li }
\affiliation{ Wuhan National High Magnetic Field Center $\&$ School of Physics, Huazhong University of Science and Technology, Wuhan 430074, China }
\author{Gang Xu}\email{gangxu@hust.edu.cn}
\affiliation{ Wuhan National High Magnetic Field Center $\&$ School of Physics, Huazhong University of Science and Technology, Wuhan 430074, China }
\date{\today}

\begin{abstract}
By means of ternary chemical potential phase diagram and phonon spectrum calculations, we propose that $M$Tl$_4$Te$_3$~($M$ = Cd, Hg), the derivatives of Tl$_5$Te$_3$, are thermodynamically and dynamically stable in the body centered tetragonal crystal structure with $I$4/$mcm$ symmetry. Our electronic structures calculations confirm that a robust $s$-$p$ band inversion occurs near the Fermi level in $M$Tl$_4$Te$_3$, and a topological band gap about 0.13~eV in CdTl$_4$Te$_3$ is induced by the spin-orbit coupling. These results suggest that $M$Tl$_4$Te$_3$ are a new class of large band gap 3D strong topological insulators that are stable and synthesizable in experiment, which could be used to design efficient spin torque equipment and spin device.

\end{abstract}

%\pacs{ }
\maketitle
%\tableofcontents

%%%%%%%%%%%%%%%%%%%%%%%%%%%%%%%%%%%%%%%%%%%%%%%%%%%%%%%%%%%%
%%BEGIN_MAIN_TEXT
%%%%%%%%%%%%%%%%%%%%%%%%%%%%%%%%%%%%%%%%%%%%%%%%%%%%%%%%%%%%

\section{Introduction}
In the past two decades, topological electronic materials, including
topological insulators (TIs)~\cite{PhysRevB.74.085308, science.1133734,
PhysRevB.78.195125, RevModPhys.82.3045, RevModPhys.83.1057}, topological
semimetals~\cite{PhysRevB.83.205101, PhysRevLett.107.186806,
PhysRevLett.107.127205, PhysRevB.85.165110, PhysRevLett.108.140405,
PhysRevB.85.195320, PhysRevB.88.125427, hsieh2012topological,
xu2012observation}, and topological
superconductors(SC)~\cite{RevModPhys.83.1057, PhysRevLett.100.096407,
PhysRevLett.102.187001, PhysRevLett.104.057001, PhysRevLett.105.097001,
PhysRevLett.117.047001, PhysRevB.81.134508}, have attracted great interest and
reshaped perception of the materials.
In particular, large band gap three-dimensional~(3D) TIs have many intriguing properties in both fundamental physics~\cite{PhysRevB.78.195424,
science.1167733, li2010dynamical} and device
applications~\cite{brumfiel2010topological, pesin2012spintronics,
APLMaterials.5.035504, PhysRevLett.121.116801,
wu2021magnetic}.
TIs with the spin momentum locked surface states are robust against perturbations,
and thus have high performance on the spin torque~\cite{Mellnik2014,RN37}, topotronic~\cite{PhysRevB.92.205310,RN35} and spintronic devices~\cite{zhang2009topological, hsieh2009tunable}.
Furthermore, many exotic topological states are realized by modulating the TIs.
For example, the quantum anomalous Hall (QAH) effect has been achieved by the magnetic doping in Bi$_2$Te$_3$ films ~\cite{PhysRevLett.61.2015, PhysRevLett.90.206601, PhysRevLett.101.146802,
AdvPhys.64.227, science.1187485,
science.1234414, xia2009observation, ge2020high}.
Besides, topological superconductivity is reported in the Cu-intercalated Bi$_2$Se$_3$ (Cu$_x$Bi$_2$Se$_3$)~\cite{PhysRevLett.104.057001,
PhysRevLett.106.127004, PhysRevLett.105.097001, wray2010observation},
Bi$_2$Te$_3$ under high pressure~\cite{Zhang24} and TI/SC
heterostructures~\cite{science.1216466}.
However, easily synthesized TIs with relatively large band gap and clear
two-dimensional~(2D) Dirac-cone surface states as the Bi$_2$Se$_3$
family~\cite{zhang2009topological, xia2009observation, hsieh2009tunable,
science.1173034, PhysRevLett.103.146401, arakane2012tunable} remain rare until now.
Therefore, it is desirable to search for large band gap 3D TIs for potential utilizations.

First-principles calculations have played remarkable roles in the development of topological physics and topological materials.
Many topological materials are predicted by first-principles calculations
firstly, and then confirmed by experiments, including HgTe
quantum well~\cite{science.1133734, science.1148047}, bismuth antimony alloy
Bi$_{1-x}$Sb$_x$\cite{PhysRevB.76.045302, hsieh2008topological,
science.1167733}, Bi$_2$Se$_3$-family of TIs~\cite{zhang2009topological,
xia2009observation, hsieh2009tunable, science.1173034, PhysRevLett.103.146401,
arakane2012tunable}, topological crystalline insulator
(TCI) SnTe~\cite{hsieh2012topological, xu2012observation}, topological
semimetals~\cite{PhysRevB.85.195320, PhysRevB.88.125427, science.1245085,
science.1256742, yi2014evidence, jeon2014landau, PhysRevLett.113.246402,
PhysRevB.83.205101, PhysRevLett.107.186806, PhysRevX.5.011029,
PhysRevX.5.031023} and so on.
Recently, Tl$_5$Te$_3$~\cite{schewe1989crystal} has been found to be a topological material that host Dirac surface states at 0.5~eV above the Fermi level~(E$_F$).
After then, by using 4$c$ site substitution,
many derivatives $M$Tl$_4$Te$_3$ ($M$ = Cu, Sn, Mo,
Pb, Bi, Sb, La, Nd, Sm, Gd, Tb, Dy, Er, Tm)~\cite{ bradtmoller1994crystal,
imamalieva2008new, PhysicsandChemistryofSolidState.21.492, RN34, bradtmoller1993darstellung}
have been experimentally synthesized and reported.
Among them, SnTl$_4$Te$_3$ with 8-electron configuration is expected to be a TI.
Unfortunately, the band inversion in SnTl$_4$Te$_3$ disappears so that it becomes a trivial
insulator~\cite{PhysRevLett.112.017002, APLMaterials.3.041507}.

Inspired by the above understanding, we propose that CdTl$_4$Te$_3$ and HgTl$_4$Te$_3$ are a new class of large band gap 3D strong topological insulators that are stable and synthesizable experimentally, through the 4$c$ site substitution of Cd or Hg. For this purpose, ternary chemical potential phase diagram with precursors and convex hull diagram are constructed. Both of them demonstrate that CdTl$_4$Te$_3$ is thermodynamically stable and easily synthesized in Cd-rich, Tl$_2$Te$_3$-rich and Tl-poor condition. The phonon spectrum reveal that CdTl$_4$Te$_3$ adopts the body centered tetragonal structure with $I$4/$mcm$ symmetry. Further electronic structures calculations identify that a robust band inversion between Cd-5$s$ and Te-5$p$ orbitals exists at $\Gamma$ point even without spin-orbit coupling~(SOC). When SOC is considered, a topological band gap about 0.13~eV is induced in CdTl$_4$Te$_3$, which is larger than the energy scale of room temperature in theory. As a result, one single Dirac cone formed by the topological surface states is discovered at $\bar{\Gamma}$ point of the surface. The corresponding left-hand momentum locking texture is also studied, which can be applied on the efficient spin torque equipment and spin device design. Finally, the topological electronic structures and stability of the other derivative HgTl$_4$Te$_3$ are discussed, which is expect to possess the same crystal structure and strong TI nature as CdTl$_4$Te$_3$.

\section{Crystals structures and methodology}
\begin{figure}{}
\includegraphics[width=\columnwidth]{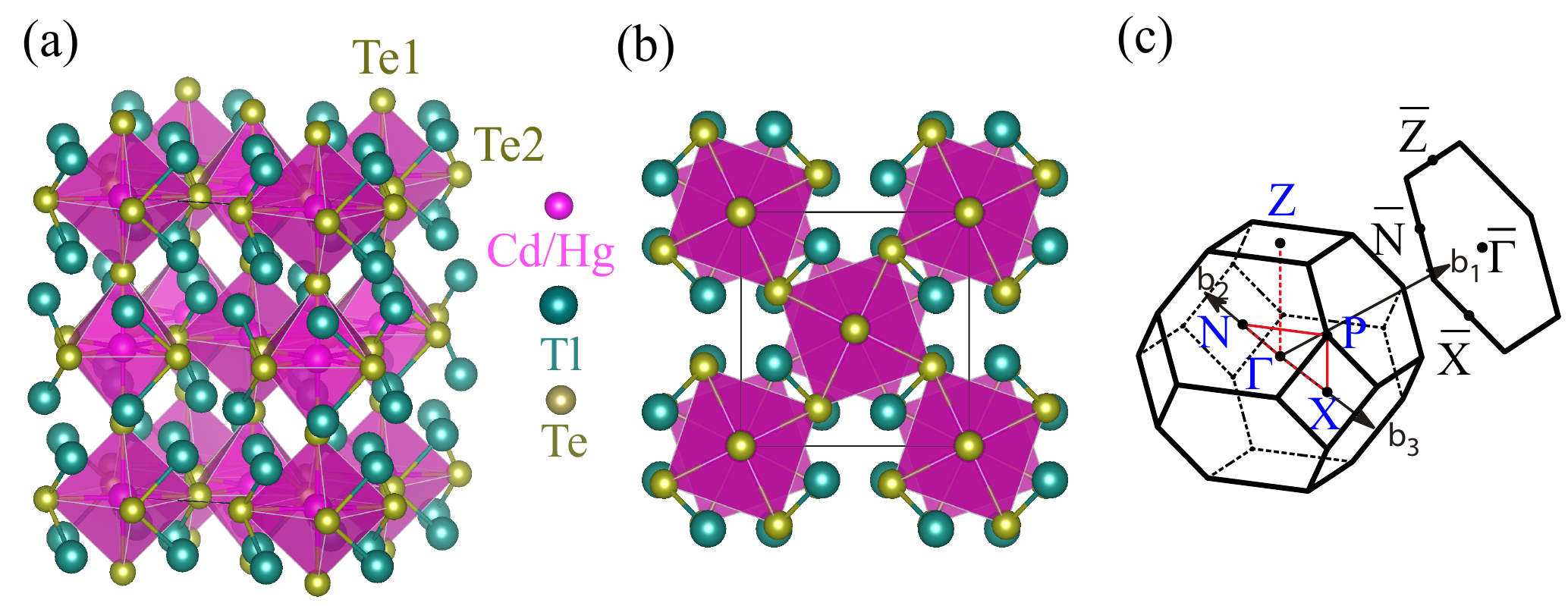}
\caption{(a) The unit cell of CdTl$_4$Te$_3$. (b) Top view of the unit cell. (c) Bulk BZ and its projection to the (100) surface of its primitive cell. High symmetry $k$ path is indicated.
\label{fig:structure}}
\end{figure}

In our work, the same crystal structure of SnTl$_4$Te$_3$, i.e., the body
centered tetragonal phase with $I4/mcm$ space group(No.~140, D$^{18}_{4h}$) as shown
in Fig.~\ref{fig:structure}(a), is used for the calculation of
$M$Tl$_4$Te$_3$~($M$ = Cd, Hg).
The corresponding first Brillouin zone~(BZ)~\cite{SETYAWAN2010299} and its projection on the (100)
surface of the primitive cell are displayed in Fig.~\ref{fig:structure}(c).
In Fig.~\ref{fig:structure}(a), $M$ atom locates at Wyckoff position 4$c$~($0.5,
~0.5, ~0.0$) which is the center of the corner-sharing CdTe$_6$ octahedron;
Tl is located at 16$l$~($x_1, ~0.5+x_1, ~z_1$);
two types of Te locate at 4$a$~($0.5, ~0.5, ~0.25$) and 8$h$~($-x_2,
~-x_2+0.5, ~0.0$).
The lattice parameters, atomic coordinates $x_1$, $x_2$ and $z_1$ are fully
relaxed.
The detailed structure information of MTl$_4$Te$_3$ is summarized in
Table.~\ref{tab:1}.
Besides, all the compounds used in our ternary phase diagram calculation are
based on their ground phase in experiment, and their crystal parameters are
fully relaxed as tabulated in Tabel.~\ref{tab:1}
to make the energies comparable.

Our first-principles calculations are performed by the Vienna ab initio
simulation package~\cite{KRESSE199615, PhysRevB.54.11169} with the projected
augmented wave method~\cite{PhysRevB.50.17953}. The energy cutoff is set as 400~eV, and $9\times9\times7$ k-meshes are adopted. Local-density approximation~(LDA) type of
the exchange-correlation potential~\cite{PhysRevB.23.5048} is used in all
calculations.
All the different compositions Cd$_l$Tl$_m$Te$_n$ are fully relaxed until the Hellmann-Feynman forces on each atom are less than 0.01 eV/\AA~and the total energy converge up to 10$^{-6}$~eV.
The ternary phase diagram is constructed by calculating the total energy of Cd$_l$Tl$_m$Te$_n$ without SOC.
The phonon spectrum calculations are carried out by the PHONOPY code~\cite{TOGO20151} with a $2\times2\times2$ supercell through the density-functional perturbation theory approach~\cite{PhysRevB.55.10355}.
The band inversion is further confirmed by the modified Becke-Johnson (MBJ) potential~\cite{PhysRevLett.102.226401} with the MBJ parameter C$_{MBJ}$ setting as 1.35. We note that C$_{MBJ}$ = 1.1 $\sim$ 1.7 is usually used for semiconductors including IIB-VIA compounds as proposed by Tran and Blaha~\cite{PhysRevLett.102.226401, PhysRevB.85.155109}. Its reliability and accuracy have been identified to be at the same level as the hybrid functional~\cite{heyd2003hybrid} and GW methods~\cite{PhysRev.139.A796} for a wide variety of semiconductors~\cite{PhysRevB.83.195134, PhysRevB.82.155145, PhysRevB.82.205212, PhysRevB.82.205102, PhysRevB.93.045304, PhysRevB.93.115104, PhysRevB.101.245163, PhysRevB.85.155109}.
The maximally localized Wannier functions are constructed by the Wannier90 package~\cite{MOSTOFI20142309} based on the MBJ+SOC calculations. The surface states are calculated by iterative
Green's function method as implemented in the WannierTools
package~\cite{WU2018405}.

\section{The stability of CdTl$_4$Te$_3$}
\begin{figure}{}
\includegraphics[width=1.0\columnwidth]{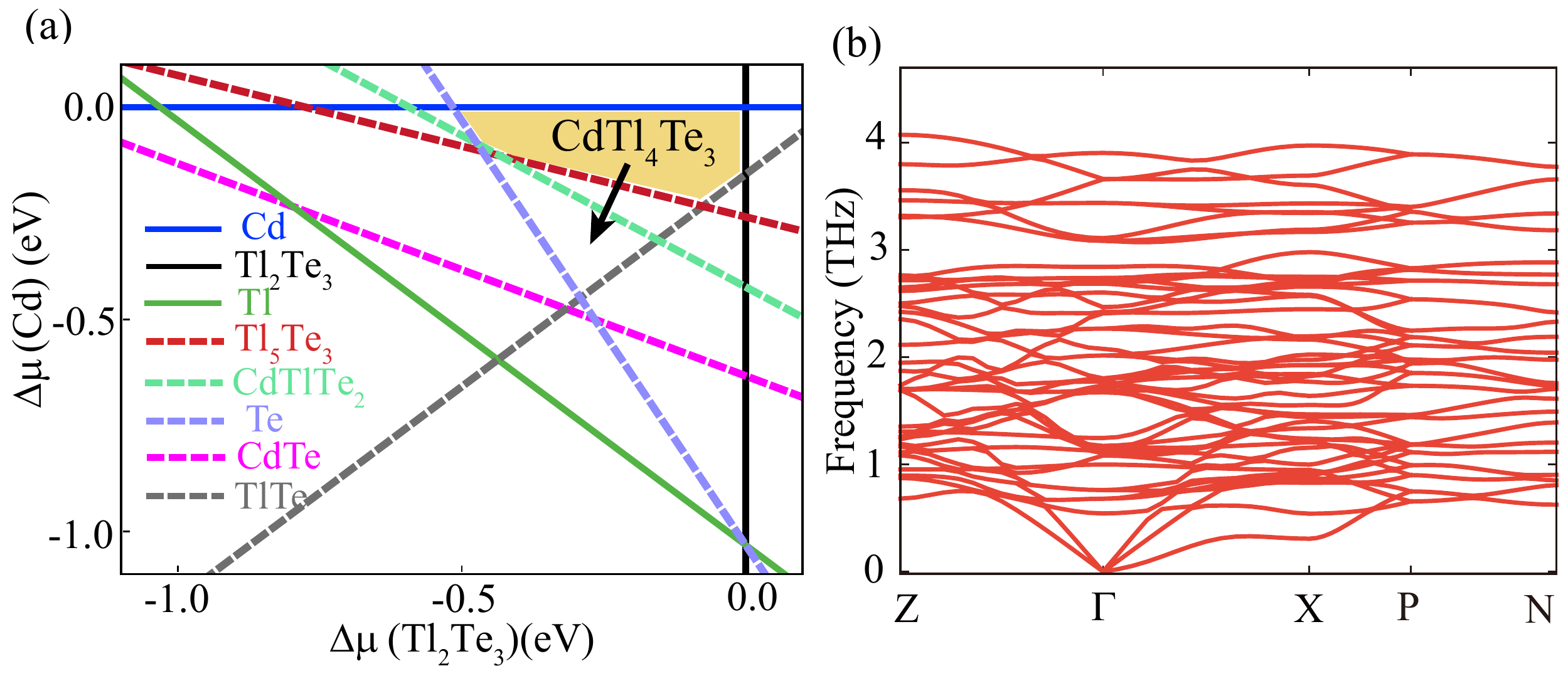}
\caption{(a) The ternary phase diagram of CdTl$_4$Te$_3$ with respect to the chemical potentials of Cd, Tl$_2$Te$_3$ and Tl, $\Delta\mu$(Tl)=$\frac{1}{2}$[E$^{F}$(CTT)-$\Delta\mu$Tl$_2$Te$_3$-$\Delta\mu$Cd].
%The grey lines represent the maximum values of $\mu$(Tl$_2$Te$_3$) and $\mu$(Cd) which is equal to formation energy of CdTl$_4$Te$_3$.
(b) The Phonon dispersion curves of CdTl$_4$Te$_3$ without SOC.
\label{fig:phase}}
\end{figure}

In this section, we would like to take CdTl$_4$Te$_3$ as example to study its thermodynamical and dynamical stability.
By choosing three stable compounds Cd,Tl and Tl$_2$Te$_3$ as the precursor materials, the target compound CdTl$_4$Te$_3$ can be synthesized by the following reaction:
\begin{equation}
Tl_2Te_3+Cd+2Tl~\rightarrow~CdTl_4Te_3.
\label{eq:syn}
\end{equation}
A phase diagram as function of the precursors' chemical potentials could determine the ranges of the experimental synthesis condition within which the target compound can be stabilized and within which the undesired competing phases are formed by varying the composition of precursors~\cite{RN36,IntegratedFerroelectrics.175.186,ma13194303,JournaloftheAmericanCeramicSociety.72.2104,RN21,RN22}. In general, the synthesis process can be understood as the exchange of elemental components between precursors and the forming phases. Therefore, the formation energies of all forming compounds Cd$_l$Tl$_m$Te$_n$ could be expressed as:
\begin{equation}
\begin{aligned}
E^{F}(Cd_lTl_mTe_n)~=~&l\Delta\mu(Cd) + \frac{n}{3}\Delta\mu(Tl_2Te_3)\\
& + (m-\frac{2n}{3})\Delta\mu(Tl).
\label{eq:formCTT1}
\end{aligned}
\end{equation}
where $\Delta\mu$ (i) = $\mu$(i)~-~E$^T$(i) with i = Cd, Tl and Tl$_2$Te$_3$ are the chemical potentials of the precursors referenced to the total energy of their ground states. Therefore, Eq.~\ref{eq:formCTT1} makes a connection between $\Delta\mu(i)$ and experimental condition, which means a rich condition of the corresponding precursor if $\Delta\mu(i)$ is close to zero, and a poor condition if $\Delta\mu(i)$ has a large negative value.
Based on our calculation, the formation energy of CdTl$_4$Te$_3$ is -1.032~eV/formula with respect to the precursors, which leads to two requirements for the chemical potentials. One is that each $\Delta\mu(i)$ could only vary between 0 and -1.032~eV. The other is that there are only two independent variable chemical potentials. Therefore, the phase diagram can be visualized by a 2D graph with variables $\Delta\mu$(Tl$_2$Te$_3$) and $\Delta\mu$(Cd) as shown in Fig.~\ref{fig:phase}(a).

The competing phases such as CdTlTe$_2$, TlTe, Tl$_5$Te$_3$, CdTe and Te are considered.
Their formation energies E$^F$(Cd$_l$Tl$_m$Te$_n$) with respect to the precursors are calculated and listed in Table~\ref{tab:2}. Then the phase diagram of CdTl$_4$Te$_3$ as function of the chemical potentials $\Delta\mu$(Cd), $\Delta\mu$(Tl$_2$Te$_3$) and $\Delta\mu$(Tl) is constructed in Fig.~\ref{fig:phase}(a) by using a general scheme~\cite{IntegratedFerroelectrics.175.186,ma13194303, RN22}. More details are described in supplementary material~(SM)~\cite{SM}.
Yielding to the constraint $\Delta\mu$(i) = 0 $\sim$ -1.032 eV, the whole allowed chemical potential region is restricted in the triangle surrounded by $\Delta\mu$(Cd)=0 (blue solid line), $\Delta\mu$(Tl)=0 (green solid line) and $\Delta\mu$(Tl$_2$Te$_3$)=0 (black solid line).
Our results reveal that CdTl$_4$Te$_3$ is most stable against to other competing compounds in the orange region manifested by the phase separation lines $\Delta\mu($Tl$_5$Te$_3$) = 0, $\Delta\mu$(CdTlTe$_2$) = 0, $\Delta\mu$(TlTe) = 0 and $\Delta\mu$(Te) = 0.
Here, $\Delta\mu$(Cd$_l$Tl$_m$Te$_n$) = $\mu$(Cd$_l$Tl$_m$Te$_n$) - E$^F$(Cd$_l$Tl$_m$Te$_n$) is the chemical potential of competing phase that is the function of precursor's chemical potentials. The competing phase Cd$_l$Tl$_m$Te$_n$ will precipitate out at $\Delta\mu$(Cd$_l$Tl$_m$Te$_n$) = 0, and become unstable with a negative value.
These results clearly demonstrate that CdTl$_4$Te$_3$ is easy to synthesize at Cd-rich, Tl$_2$Te$_3$-rich and Tl-poor condition.
We would like to recall that the abundance of precursors is relative yielding to E$^F$(CdTl$_4$Te$_3$)~=~-1.032~eV and Eq.~\ref{eq:formCTT1}.
By increasing $\Delta\mu$(Tl) along the black arrow in Fig.~\ref{fig:phase}(a), it means that Tl grows more and more rich while Cd and Tl$_2$Te$_3$ become poor.
When the arrow cross $\Delta\mu$(Tl$_5$Te$_3$)=0 and $\Delta\mu$(CdTlTe$_2$)=0 lines, CdTl$_4$Te$_3$ becomes unstable accompanied with the precipitation of Tl$_5$Te$_3$ and CdTlTe$_2$, and the following decomposition will take place,
\begin{equation}
\label{eq:decompose}
\begin{split}
CdTl_4Te_3~&\rightarrow~Tl_5Te_3+3TlTe+2Cd\\
CdTl_4Te_3~&\rightarrow~CdTlTe_2+TlTe+2Tl\\
\end{split}
\end{equation}
\newline
According to our calculations, the energy of the right products is 0.234~eV and 0.564~eV higher than that of left, respectively.

The convex hull analysis is another useful method to investigate the thermodynamical stability~\cite{JChemPhys.154.234706, PhysRevLett.123.097001}.
In this way, the formation energy of all possible atomic configurations with respect to constituent elements needs to be calculated, which is defined as $\widetilde{E}^F = E^T(Cd_lTl_mTe_n) - lE^T(Cd) - mE^T(Tl) - nE^T(Te)$.
Using this definition, our calculations demonstrate that CdTl$_4$Te$_3$ is thermodynamically stable against to elements with formation energy of -0.213~eV/atom.
With all possible $\widetilde{E}^F$ as listed in Table.~\ref{tab:2}, the convex hull diagram is constructed in Fig.~S1~\cite{SM}, which shows that only the binary compounds are on the convex hull, and all ternary compounds such as CdTl$_4$Te$_3$ and CdTlTe$_2$ are within a viable energy window for potentially metastable phases.
Even so, we estimate that CdTl$_4$Te$_3$ is potentially synthesizable based on the following facts.
One is that CdTl$_4$Te$_3$ is just a little above the convex hull with small energy and even 0.020 eV/atom lower than CdTlTe$_2$.
Since CdTlTe$_2$ has already been synthesized in 1969~\cite{GUSEINOV1969807}, it is expected that CdTl$_4$Te$_3$ is highly feasible in proper condition, especially at Cd-rich, Tl$_2$Te$_3$-rich and Tl-poor condition in the reaction Eq.~\ref{eq:formCTT1}.

\begin{table}[htbp]\footnotesize
\centering
\setlength{\tabcolsep}{0.3mm}
\caption{Detailed crystallographic information of corresponding compounds used in Fig.~\ref{fig:phase}.}
\begin{tabular}{cccccccccc}
\toprule[0.8pt]
Compounds & Space group &{$a, b, c$}&&Atomic coordinates \\
& &\AA & &(fractional)\\
\midrule[0.3pt]
Cd & $P6_3/mmc$ & $a$=2.9179 & Cd & $2$c $ (\frac{1}{3}$  $\frac{2}{3}$ 0.25) &  \\
& &$c$=5.3842 &&\\
Tl & $P6_3/mmc$ & $a$=3.4108 & Tl& 2$c$ ($\frac{1}{3}$  $\frac{2}{3}$  $0.25$) &  \\
& &$c$=5.4269 &&\\
Te & $P3_121$ & $a$=4.2786 & Te & 3$a$ ($0.2879$  $0$  $\frac{1}{3}$) &  \\
& &$c$=5.9249 &&\\
CdTe & $F-43m$ & $a$=6.4082 & Cd & 4$a$ ($0.0$  $0.0$  $0$) &  \\
& & &Te &4$c$ (0.25 0.25 0.25)&\\
TlTe & $I4/mcm$ & $a$=12.6445 & Tl & 16k (0.0783 0.2300 0.0) &  \\
& &$c=6.0775$ &Te1 &8$h$ (0.1663 0.6663 0.25)&\\
& & &Te2 &4$d$ (0.0 0.5 0.0)&\\
& & &Te3 &4$a$ (0.0 0.0 0.25)&\\
Tl$_2$Te$_3$ & $C2/c$ & $a$=12.8449 & Tl1 & 8$f$ (0.8947 0.6452 0.5521) &  \\
& $\alpha=\gamma=90^{\circ}$&$b$=6.3989 &Te1 &8$f$ (0.6817 0.6379 0.0860)&\\
& $\beta=144.9264^{\circ}$&$c$=13.0071 &Te2 &4$e$ (0.0 0.8703 0.25)&\\
CdTlTe$_2$ & $P-3m1$ & $a$=4.2425 & Cd & 1$a$ (0.0 0.0 0.0) &  \\
& &$c$=7.3925 &Tl &1$b$ (0.0 0.0 0.5)&\\
& & &Te &2$d$ ($\frac{1}{3}$ $\frac{2}{3}$ 0.2236)&\\
Tl$_5$Te$_3$ & $I4/mcm$ & $a$=8.6689 & Tl1 & 4$c$ (0.5 0.5 0.0) &  \\
& &c=12.6155 &Tl2 &16$l$ (0.1476 0.6476 0.1607)&\\
& & &Te1 &4$a$ (0.5 0.5 0.25)&\\
& & &Te2 &8$h$ (0.6565 0.1565 0.0)&\\
CdTl$_4$Te$_3$ & $I4/mcm$ & a=8.5120 & Cd & 4$c$ (0.5 0.5 0.0) &  \\
& &$c$=12.2642 &Tl &16$l$ (0.1452 0.6452 0.1651)&\\
& & &Te1 &4$a$ (0.5 0.5 0.25)&\\
& & &Te2 &8$h$ (0.6624 0.1624 0.0)&\\
HgTl$_4$Te$_3$ & $I4/mcm$ & a=8.6668 & Hg & 4$c$ (0.5 0.5 0.0) &  \\
& &$c$=12.1854 &Tl &16$l$ (0.1460 0.6460 0.1624)&\\
 & & &Te1 &4$a$ (0.5 0.5 0.25)&\\
& & &Te2 &8$h$ (0.6599 0.1599 0.0)&\\
\bottomrule[0.8pt]
\end{tabular}
\label{tab:1}
\end{table}

\begin{table}[htbp]
\centering
\caption{The calculated energy/formula used in Fig.~\ref{fig:phase}.}
%$\widetilde{E}^F = E^T(Cd_lTl_mTe_n) - lE^T(Cd) - mE^T(Tl) - nE^T(Te)$.
%%\newcommand{\tabincell}[2]{\begin{tabular}{@{}#1@{}}#2\end{tabular}}
\begin{tabular}{ccccc}
\toprule[0.8pt]
\makebox[0.1\textwidth]{Compounds} & \makebox[0.2\textwidth]{E$^T$(eV)} & \makebox[0.08\textwidth] {E$^F$(eV)}  & \makebox[0.08\textwidth]{$\widetilde{\rm{E}}^F$(eV)} & \\
\midrule[0.3pt]
Cd & -1.5037 & \ & \ &  \\
\midrule[0.3pt]
%\arrayrulecolor{gray} \hdashline{2-3}[0.8pt/2pt]
%\hdashline
Tl & -2.9543 & \ & \ & \\
\midrule[0.3pt]
Te & -3.8046 & 0.2237 & \ & \\
\midrule[0.3pt]
Tl$_2$Te$_3$ & -17.9934 & \ &-0.6709 &  \\
\midrule[0.3pt]
CdTe & -6.0324 & -0.5004 & -0.7241 &  \\
\midrule[0.3pt]
TlTe & -7.1281 & -0.1455 & -0.3691 &  \\
\midrule[0.3pt]
Tl$_5$Te$_3$ & -28.0154 & -1.1591 & -1.8300 &  \\
\midrule[0.3pt]
CdTlTe$_2$ & -12.8369 & -0.3224 & -0.7697 &  \\
\midrule[0.3pt]
CdTl$_4$Te$_3$ & -26.4372 & -1.0315 & -1.7025 & \\
\bottomrule[0.8pt]
\end{tabular}
\label{tab:2}
\end{table}

To further check the dynamical stability of CdTl$_4$Te$_3$, the phonon spectrum based on the body centered tetragonal phase with $I$4/$mcm$ symmetry is calculated. As shown in Fig.~\ref{fig:phase}(b), there is no phonon mode with negative frequency in the entire BZ, which indicates that CdTl$_4$Te$_3$ is dynamically stable by adopting the body centered tetragonal structure. The above thermodynamical and dynamical investigations strongly demonstrate that CdTl$_4$Te$_3$ is readily synthesized in experiment.

\section{Electronic properties of CdTl$_4$Te$_3$}
\begin{figure}{}
\includegraphics[width=\columnwidth]{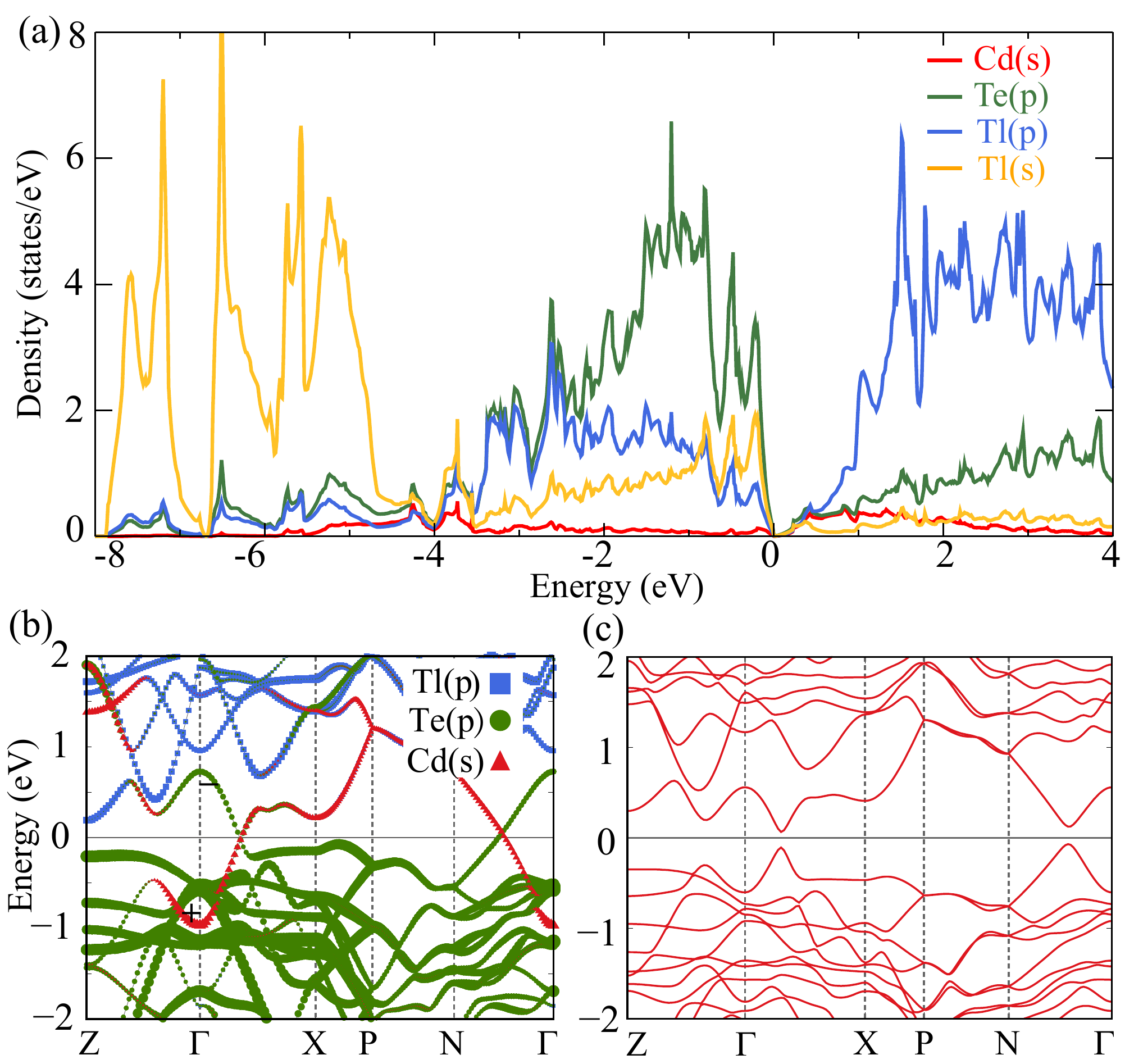}
\caption{ The electronic properties of CdTl$_4$Te$_3$. (a) The projected DOS without SOC.
(b) The fatted bands with spectral weight of Cd-5$s$(red), Te-5$p$(green) and Tl-6$p$(blue) orbitals without SOC. (c) The band structures of with MBJ by including SOC.
\label{fig:band}}
\end{figure}

The projected density of states (PDOS) of tetragonal CdTl$_4$Te$_3$ are calculated and plotted. As shown in Fig.~\ref{fig:band}(a), Tl-6s orbitals mainly contribute to the states between -8~$\sim$~-3.5~eV, while Tl-6p orbitals mainly contribute to the states above 0.3~eV.
Considering the electronic configuration $6s^26p^1$ of Tl, we conclude that Tl
favors +1 valence in CdTl$_4$Te$_3$, similar to valence of
Tl atoms at 16$l$ site in Tl$_5$Te$_3$~\cite{APLMaterials.3.041507, PhysicsandChemistryofSolidState.21.492}.
The states in energy range of -3.5~$\sim$~0~eV are approximately from Te-5$p$
orbitals with admixing of Tl-6$p$ and Cd-5$s$ states,
which implies that the bonding between Cd, Tl and Te is not the pure ionic bond but has sizable metal-metal bond characters.
Fig.~\ref{fig:band}(a) shows that the Cd-5$s$ orbitals are almost empty and mainly contribute the states between 0~$\sim$~3~eV.
Therefore, we can understand the electron transfer roughly as
follows. Each Tl atom donates one 6$p$ electron, and Cd atom donates two 5$s$
electrons to the Te-5$p$ orbitals.
As a result, CdTl$_4$Te$_3$ is close to the electronic configuration of the atomic insulator with the full filled sub-shell of Te$^{2-}$, Tl$^{1+}$, Cd$^{2+}$ ions approximately.

However, we notice that the Cd-5$s$ orbitals are very extended,
which also exhibit considerable amplitude under the E$_F$, implying a band inversion between Cd-5$s$ and Te-5$p$ states.
The character is further verified by the projected band structures in
Fig.~\ref{fig:band}(b), which clearly demonstrates that
the Cd-5$s$ states with even parity are lower 1.82~eV
than Te-5$p$ states with odd parity at $\Gamma$ point.
The band inversion in CdTl$_4$Te$_3$ is already occurred even without SOC and can
be alternatively viewed as a consequence of the inert pair effect in chemistry,
which is the propensity for the two electrons in the outermost 5$s$ orbital to
remain unionized in heavier elements~\cite{narang2021topology}, just like that in HgTe~\cite{science.1133734, PhysRevB.74.085308}.
In Fig.~\ref{fig:band}(b), when SOC is excluded, the band crossing points between Cd-5$s$ and Te-5$p$
can be protected by time reversal
symmetry~($\mathcal{TRS}$) and inversion symmetry~($\mathcal{I}$) symmetry and
form nodal rings as plotted in Fig.~S2(a)~\cite{SM}.
Since LDA type exchange-correlation potential usually overestimates band inversion between valence and conduction bands, then MBJ~\cite{PhysRevLett.102.226401} is employed.
The amplitude of band inversion in LDA is reduced to 1.10~eV with MBJ calculations as shown in Fig.~S2(b)~\cite{SM}.
Therefore, the band inversion in CdTl$_4$Te$_3$ is robust against the functional potentials, and the more accurate calculation of LDA with MBJ semilocal exchange
functional potential is adopted to investigate the electronic and topological properties in the following.

When SOC is considered, the nodal rings are all gapped, inducing a 0.13~eV band gap
as shown in Fig.~\ref{fig:band}(c), which is larger than the energy scale of
room temperature theoretically.
For insulators with $\mathcal{I}$, the topological invariant $\nu_0$
based on the Fu-Kane formula~\cite{PhysRevLett.98.106803}
can be characterized by the parity products~($\xi_i$) of the half numbers of the occupied states at eight time-reversal invariant
momentum~(TRIM) points~(Kramers pairs have the same parities).
As shown in Fig.~\ref{fig:structure}(b), there are one $\Gamma$~(0.0, ~0.0,
~0.0), one $Z$~(0.5, ~0.5, ~-0.5), two $X$~(0.0, ~0.0, ~0.5) and four $N$~(0.5,
~0.0, ~0.0) TRIM points in the first BZ.
Therefore, only $\xi_{\Gamma}$ and $\xi_Z$ could determine the topological property of the tetragonal CdTl$_4$Te$_3$,
while the other TRIM points always give the trivial products.
Our calculations indicate $\xi_{\Gamma}=-1$ and $\xi_Z=1$, and give rise to $\nu_0=1$. These results are consistent with the band inversion analysis at $\Gamma$ point, and confirm that CdTl$_4$Te$_3$ is a strong TI.

\begin{figure}[h]
\includegraphics[width=\columnwidth]{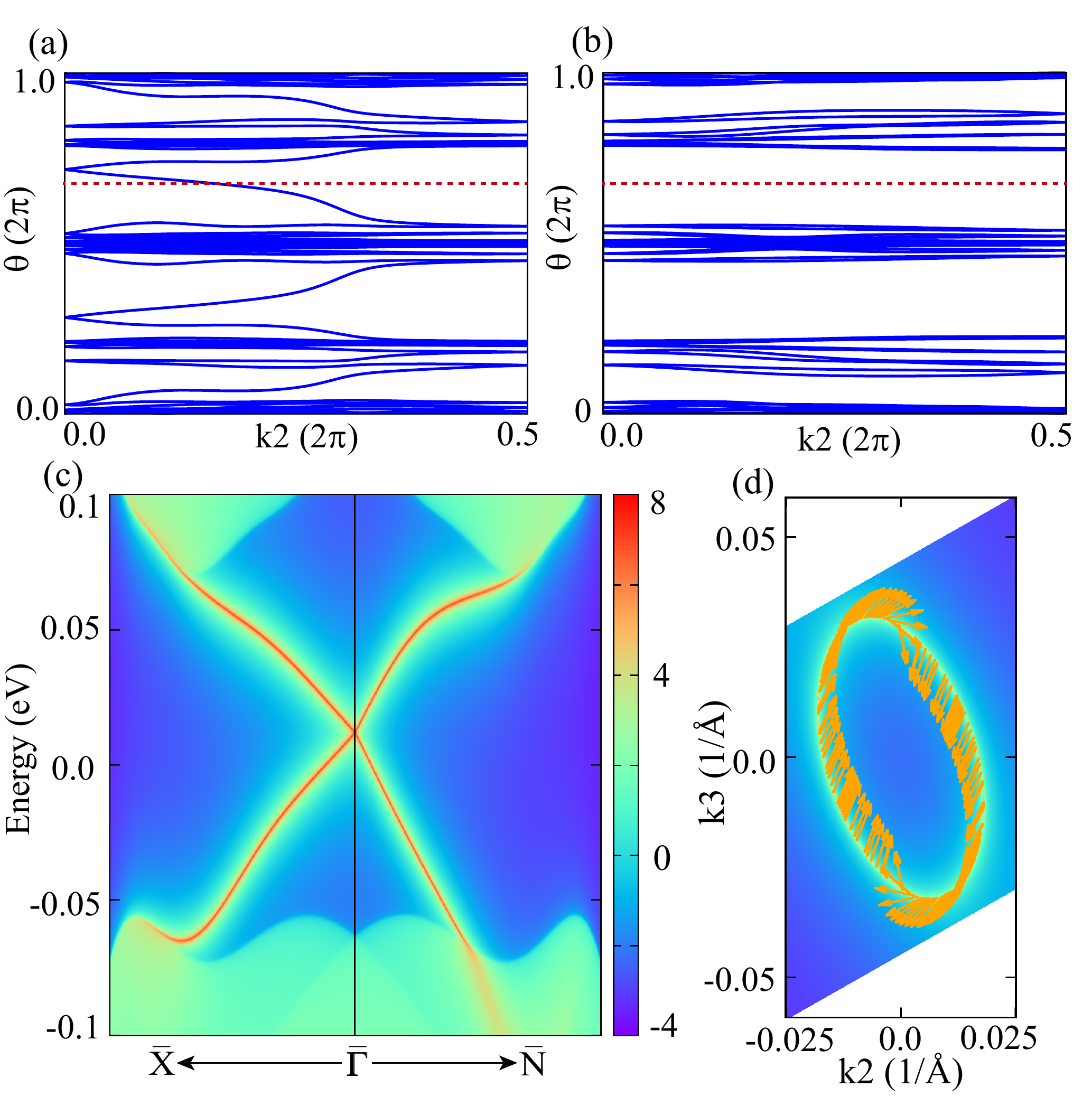}
\caption{(a) The Wilson loops of k$_2$k$_3$ plane (k$_1$ = 0) and (b) (k$_1$ = $\pi$) plane. (c) Band structures projected onto the (100) surface. (d) Topological surface states with chemical potential at 0.02~eV and corresponding spin texture on the (100) surface.
\label{fig:surface}}
\end{figure}

\section{Topological properties of CdTl$_4$Te$_3$}
We construct the maximally localized Wannier functions of Cd-5$s$, Te-5$p$ and Tl-6$p$ states to
investigate the topological features more explicitly.
The Wilson loop method~\cite{PhysRevB.84.075119} is used by calculating the evolution of Wannier charge
centers for the occupied bands in k$_1$ = 0~(Fig.~\ref{fig:surface}(a)) and k$_1 = \pi$ planes~(Fig.~\ref{fig:surface}(b)).
The evolution lines cross the reference line(red dashed line) once time in Fig.~\ref{fig:surface}(a), indicating that the k$_1$ = 0 plane corresponds to a quantum spin Hall system with non-trivial 2D topological invariant. The evolution lines cross the reference line(red dashed line) zero time in Fig.~\ref{fig:surface}(b), confirming that the k$_1$ = $\pi$ plane is a trivial 2D system. These results combining with Wilson loops on other surface planes(Fig.~S3)~\cite{SM}, give rise a complete topological index $\mathbb Z_2$=(1;000), which further confirm that CdTl$_4$Te$_3$ fall into the strong TI phase.
In Fig.~\ref{fig:surface}(c), we plot the surface electronic structures on the (100) surface by the iterative surface Green's function method~\cite{JPhysFMetPhys.14.1205, JPhysFMetPhys.15.851}. Two robust surface states connect the valence and conduction bands and form a Dirac cone in the bulk gap at $\bar{\Gamma}$ point due to the requirement of the $\mathcal{TRS}$.
In Fig.~\ref{fig:surface}(d), we plot the Fermi surfaces of the Dirac cone at 0.02~eV and their spin orientation, which exhibits a left-hand spin texture enclosed a $\pi$ phase like that in Bi$_2$Se$_3$~\cite{zhang2009topological}, indicating a positive SOC in CdTl$_4$Te$_3$~\cite{Nie10596}. Such kind of spin momentum locking surface states have been reported that have very highly efficient performance on spin torque equipment~\cite{Mellnik2014, RN37} and spin device~\cite{zhang2009topological, hsieh2009tunable}.

\section{Discussion and Conclusion}
Considering that Hg is isoelectronic with Cd in the \uppercase\expandafter{\romannumeral2}B group of the periodic table, HgTl$_4$Te$_3$ is naturally expected to be stabilized into the same crystal structure of CdTl$_4$Te$_3$ with similar electronic structures and topological property.
By using the optimized structure parameters in Table.~\ref{tab:1}, the calculated formation energy of HgTl$_4$Te$_3$ through the reaction Hg + 4Tl + 3Te~$\rightarrow$~HgTl$_4$Te$_3$ is -1.459~eV with respect to elemental precursors.
Limited by the stoichiometry, the products $\frac{1}{2}$Tl$_5$Te$_3$ + HgTe + $\frac{1}{2}$TlTe + Tl, Tl$_2$Te$_3$ + Hg + 2Tl, 3TlTe + Hg + Tl and 2TlTe + HgTe + 2Tl are taken into account by calculating the formation energy which are equal to -1.397, -0.671, -1.107, and -1.035~eV, respectively.
Obviously, HgTl$_4$Te$_3$ will be formed in the reaction because it is thermodynamically favourable.
The MBJ calculated band structures are presented in Fig.~S4~\cite{SM}.
As expected, the Hg-5$s$ state with even parity is lower 2.26~eV than Te-5$p$ state with odd parity at $\Gamma$ point.
Furthermore, 0.046~eV nontrivial band gap induced by SOC makes it a strong TI which is similar to CdTl$_4$Te$_3$.

In conclusion, we predict a new class of 3D topological insulators $M$Tl$_4$Te$_3$~($M$ = Cd, Hg) by using first-principles calculations. Our ternary chemical potential phase diagrams and phonon spectrum calculations demonstrate that $M$Tl$_4$Te$_3$~($M$ = Cd, Hg) are both thermodynamically and dynamically stable in the body centered tetragonal crystal structure with $I$4/$mcm$ symmetry.
Further electronic structures calculations confirm that the nontrivial band topology stems from the band inversion between $M$-5$s$ and Te-5$p$ orbitals at $\Gamma$ point,
and the SOC induced topological band gap is about 0.13~eV in CdTl$_4$Te$_3$ which is larger than the energy scale of room temperature in theory. The isolated Dirac cone type surface states with left-hand helicity of the spin momentum locking texture is obtained in the (100) surface spectra at $\bar{\Gamma}$ point.
These results suggest that $M$Tl$_4$Te$_3$ are synthesizable and suitable for the study of efficient spin torque equipment and spin device, which should stimulate many experimental efforts in the future.

\section{Acknowledgments}This work was supported by the National Key Research
and Development Program of China (2018YFA0307000),
and the National Natural Science Foundation of China (11874022).


\begin{thebibliography}{100}
%%No. 1    %Ref. 9  129-142
\bibitem{PhysRevB.74.085308}
X.-L. Qi, Y.-S. Wu, and S.-C. Zhang,
{\emph{Topological Quantization of the Spin Hall Effect in Two-Dimensional Paramagnetic Semiconductors}},
\href{https://link.aps.org/doi/10.1103/PhysRevB.74.085308}
{{Phys. Rev. B} {\bf{74}}, 085308 (2006)}.

%%No. 2    %Ref. 8  116-126
\bibitem{science.1133734}
B. A. Bernevig, T. L. Hughes, and S.-C. Zhang,
{\emph{Quantum Spin Hall Effect and Topological Phase Transition in $\mathrm{HgTe}$ Quantum Wells}},
\href{https://www.science.org/doi/abs/10.1126/science.1133734}
{{Science} {\bf{314}}, 1757 (2006)}.

%%No. 3    %Ref. 29  430-443
\bibitem{PhysRevB.78.195125}
A. P. Schnyder, S. Ryu, A. Furusaki, and A. W. W. Ludwig,
{\emph{Classification of Topological Insulators and Superconductors in Three Spatial Dimensions}},
\href{https://link.aps.org/doi/10.1103/PhysRevB.78.195125}
{{Phys. Rev. B} {\bf{78}}, 195125 (2008)}.

%%No. 4    %Ref. 15  220-233
\bibitem{RevModPhys.82.3045}
M. Z. Hasan and C. L. Kane,
{\emph{: Topological Insulators}},
\href{https://link.aps.org/doi/10.1103/RevModPhys.82.3045}
{{Rev. Mod. Phys.} {\bf{82}}, 3045 (2010)}.

%%No. 5    %Ref. 16  236-249
\bibitem{RevModPhys.83.1057}
X.-L. Qi and S.-C. Zhang,
{\emph{Topological Insulators and Superconductors}},
\href{https://link.aps.org/doi/10.1103/RevModPhys.83.1057}
{{Rev. Mod. Phys.} {\bf{83}}, 1057 (2011)}.

%%No. 6    %Ref. 55  815-828
\bibitem{PhysRevB.83.205101}
X. Wan, A. M. Turner, A. Vishwanath, and S. Y. Savrasov,
{\emph{Topological Semimetal and Fermi-Arc Surface States in the Electronic Structure of Pyrochlore Iridates}},
\href{https://link.aps.org/doi/10.1103/PhysRevB.83.205101}
{{Phys. Rev. B} {\bf{83}}, 205101 (2011)}.

%%No. 7    %Ref. 56  831-844
\bibitem{PhysRevLett.107.186806}
G. Xu, H. Weng, Z. Wang, X. Dai, and Z. Fang,
{\emph{Chern Semimetal and the Quantized Anomalous Hall Effect in ${\mathrm{HgCr}}_{2}{\mathrm{Se}}_{4}$}},
\href{https://link.aps.org/doi/10.1103/PhysRevLett.107.186806}
{{Phys. Rev. Lett.} {\bf{107}}, 186806 (2011)}.

%%No. 8    %Ref. 84  1239-1252
\bibitem{PhysRevLett.107.127205}
A. A. Burkov and L. Balents,
{\emph{Weyl Semimetal in a Topological Insulator Multilayer}},
\href{https://link.aps.org/doi/10.1103/PhysRevLett.107.127205}
{{Phys. Rev. Lett.} {\bf{107}}, 127205 (2011)}.

%%No. 9    %Ref. 85  1255-1268
\bibitem{PhysRevB.85.165110}
A. A. Zyuzin, S. Wu, and A. A. Burkov,
{\emph{Weyl Semimetal with Broken Time Reversal and Inversion Symmetries}},
\href{https://link.aps.org/doi/10.1103/PhysRevB.85.165110}
{{Phys. Rev. B} {\bf{85}}, 165110 (2012)}.

%%No. 10    %Ref. 86  1271-1284
\bibitem{PhysRevLett.108.140405}
S. M. Young, S. Zaheer, J. C. Y. Teo, C. L. Kane, E. J. Mele, and A. M. Rappe,
{\emph{Dirac Semimetal in Three Dimensions}},
\href{https://link.aps.org/doi/10.1103/PhysRevLett.108.140405}
{{Phys. Rev. Lett.} {\bf{108}}, 140405 (2012)}.

%%No. 11    %Ref. 48  713-726
\bibitem{PhysRevB.85.195320}
Z. Wang, Y. Sun, X.-Q. Chen, C. Franchini, G. Xu, H. Weng, X. Dai, and Z. Fang,
{\emph{Dirac Semimetal and Topological Phase Transitions in ${A}_{3}\mathrm{Bi}$ ($A=\mathrm{Na}, \mathrm{K}, \mathrm{Rb}$)}},
\href{https://link.aps.org/doi/10.1103/PhysRevB.85.195320}
{{Phys. Rev. B} {\bf{85}}, 195320 (2012)}.

%%No. 12    %Ref. 49  729-742
\bibitem{PhysRevB.88.125427}
Z. Wang, H. Weng, Q. Wu, X. Dai, and Z. Fang,
{\emph{Three-Dimensional Dirac Semimetal and Quantum Transport in $\mathrm{Cd}_3\mathrm{As}_2$}},
\href{https://link.aps.org/doi/10.1103/PhysRevB.88.125427}
{{Phys. Rev. B} {\bf{88}}, 125427 (2013)}.

%%No. 13    %Ref. 21  309-319
\bibitem{hsieh2012topological}
T. H. Hsieh, H. Lin, J. Liu, W. Duan, A. Bansil, and L. Fu,
{\emph{Topological Crystalline Insulators in the $\mathrm{SnTe}$ Material Class}},
\href{https://doi.org/10.1038/ncomms1969}
{{Nat. Commun.} {\bf{3}}, 1 (2012)}.

%%No. 14    %Ref. 22  322-334
\bibitem{xu2012observation}
S.-Y. Xu, C. Liu, N. Alidoust, M. Neupane, D. Qian, I. Belopolski, J. D. Denlinger, Y. J. Wang, H. Lin, L. A. Wray, G. Landolt, B. Slomski, J. H. Dil, A. Marcinkova, E. Morosan, Q. Gibson, R. Sankar, F. C. Chou, R. J. Cava, A. Bansil, and M. Z. Hasan,
{\emph{Observation of a Topological Crystalline Insulator Phase and Topological Phase Transition in $\mathrm{Pb}_{1\ensuremath{-}x}\mathrm{Sn}_{x}\mathrm{Te}$}},
\href{https://doi.org/10.1038/ncomms2191}
{{Nat. Commun.} {\bf{3}}, 1192 (2012)}.

%%No. 15    %Ref. 87  1287-1300
\bibitem{PhysRevLett.100.096407}
L. Fu and C. L. Kane,
{\emph{Superconducting Proximity Effect and Majorana Fermions at the Surface of a Topological Insulator}},
\href{https://link.aps.org/doi/10.1103/PhysRevLett.100.096407}
{{Phys. Rev. Lett.} {\bf{100}}, 096407 (2008)}.

%%No. 16    %Ref. 28  414-427
\bibitem{PhysRevLett.102.187001}
X.-L. Qi, T. L. Hughes, S. Raghu, and S.-C. Zhang,
{\emph{Time-Reversal-Invariant Topological Superconductors and Superfluids in Two and Three Dimensions}},
\href{https://link.aps.org/doi/10.1103/PhysRevLett.102.187001}
{{Phys. Rev. Lett.} {\bf{102}}, 187001 (2009)}.

%%No. 17    %Ref. 39  579-592
\bibitem{PhysRevLett.104.057001}
Y. S. Hor, A. J. Williams, J. G. Checkelsky, P. Roushan, J. Seo, Q. Xu, H. W. Zandbergen, A. Yazdani, N. P. Ong, and R. J. Cava,
{\emph{Superconductivity in ${\mathrm{Cu}}_{x}{\mathrm{Bi}}_{2}{\mathrm{Se}}_{3}$ and its Implications for Pairing in the Undoped Topological Insulator}},
\href{https://link.aps.org/doi/10.1103/PhysRevLett.104.057001}
{{Phys. Rev. Lett.} {\bf{104}}, 057001 (2010)}.

%%No. 18    %Ref. 88  1303-1316
\bibitem{PhysRevLett.105.097001}
L. Fu and E. Berg,
{\emph{Odd-Parity Topological Superconductors: Theory and Application to ${\mathrm{Cu}}_{x}{\mathrm{Bi}}_{2}{\mathrm{Se}}_{3}$}},
\href{https://link.aps.org/doi/10.1103/PhysRevLett.105.097001}
{{Phys. Rev. Lett.} {\bf{105}}, 097001 (2010)}.

%%No. 19    %Ref. 11  161-174
\bibitem{PhysRevLett.117.047001}
G. Xu, B. Lian, P. Tang, X.-L. Qi, and S.-C. Zhang,
{\emph{Topological Superconductivity on the Surface of $\mathrm{Fe}$-Based Superconductors}},
\href{https://link.aps.org/doi/10.1103/PhysRevLett.117.047001}
{{Phys. Rev. Lett.} {\bf{117}}, 047001 (2016)}.

%%No. 20    %Ref. 30  446-459
\bibitem{PhysRevB.81.134508}
X.-L. Qi, T. L. Hughes, and S.-C. Zhang,
{\emph{Topological Invariants for the Fermi Surface of a Time-Reversal-Invariant Superconductor}},
\href{https://link.aps.org/doi/10.1103/PhysRevB.81.134508}
{{Phys. Rev. B} {\bf{81}}, 134508 (2010)}.



%%No. 21    %Ref. 5  70-83
\bibitem{PhysRevB.78.195424}
X.-L. Qi, T. L. Hughes, and S.-C. Zhang,
{\emph{Topological Field Theory of Time-Reversal Invariant Insulators}},
\href{https://link.aps.org/doi/10.1103/PhysRevB.78.195424}
{{Phys. Rev. B} {\bf{78}}, 195424 (2008)}.

%%No. 22    %Ref. 13  192-202
\bibitem{science.1167733}
D. Hsieh, Y. Xia, L. Wray, D. Qian, A. Pal, J. H. Dil, J. Osterwalder, F. Meier, G. Bihlmayer, C. L. Kane, Y. S. Hor, R. J. Cava, and M. Z. Hasan,
{\emph{Observation of Unconventional Quantum Spin Textures in Topological Insulators}},
\href{https://www.science.org/doi/abs/10.1126/science.1167733}
{{Science} {\bf{323}}, 919 (2009)}.

%%No. 23    %Ref. 6  86-98
\bibitem{li2010dynamical}
R. Li, J. Wang, X.-L. Qi, and S.-C. Zhang,
{\emph{Dynamical Axion Field in Topological Magnetic Insulators}},
\href{https://doi.org/10.1038/nphys1534}
{{Nat. Phys.} {\bf{6}}, 284 (2010)}.



%%No. 24    %Ref. 43  639-651
\bibitem{brumfiel2010topological}
G. Brumfiel,
{\emph{Topological Insulators: Star Material}},
\href{https://doi.org/10.1038/466310a}
{{Nature} {\bf{466}}, 310 (2010)}.

%%No. 25    %Ref. 79  1167-1179
\bibitem{pesin2012spintronics}
D. Pesin and A. H. MacDonald,
{\emph{Spintronics and Pseudospintronics in Graphene and Topological Insulators}},
\href{https://doi.org/10.1038/nmat3305}
{{Nat. Mater.} {\bf{11}}, 409 (2012)}.

%%No. 27    %Ref. 40  595-605
\bibitem{APLMaterials.5.035504}
A. Politano, L. Viti, and M. S. Vitiello,
{\emph{Optoelectronic Devices, Plasmonics, and Photonics with Topological Insulators}},
\href{https://doi.org/10.1063/1.4977782}
{{APL Materials} {\bf{5}}, 035504 (2017)}.

%%No. 28    %Ref. 42  623-636
\bibitem{PhysRevLett.121.116801}
M. Ezawa,
{\emph{Topological Switch Between Second-Order Topological Insulators and Topological Crystalline Insulators}},
\href{https://link.aps.org/doi/10.1103/PhysRevLett.121.116801}
{{Phys. Rev. Lett.} {\bf{121}}, 116801 (2018)}.

%%No. 29    %Ref. 41  608-620
\bibitem{wu2021magnetic}
H. Wu, A. Chen, P. Zhang, H. He, J. Nance, C. Guo, J. Sasaki, T. Shirokura, P. N. Hai, B. Fang, S. A. Razavi, K. Wong, Y. Wen, Y. Ma, G. Yu, G. P. Carman, X. Han, X. Zhang, and K. L. Wang,
{\emph{Magnetic Memory Driven by Topological Insulators}},
\href{https://doi.org/10.1038/s41467-021-26478-3}
{{Nat. Commun.} {\bf{12}}, 6251 (2021)}.



%%NO. 30
\bibitem{Mellnik2014}
A. R. Mellnik, J. S. Lee, A. Richardella, J. L. Grab, P. J. Mintun, M. H. Fischer, A. Vaezi, A. Manchon, E. A. Kim, N. Samarth, and D. C. Ralph,
{\emph{Spin-transfer torque generated by a topological insulator}},
\href{https://doi.org/10.1038/nature13534}
{Nature {\bf 511}, 449 (2014)}.

%%NO. 31
\bibitem{RN37}
K. Kondou, R. Yoshimi, A. Tsukazaki, Y. Fukuma, J. Matsuno, K. S. Takahashi, M. Kawasaki, Y. Tokura, and Y. Otani,
{\emph{Fermi-level-dependent charge-to-spin current conversion by Dirac surface states of topological insulators}},
\href{https://doi.org/10.1038/nphys3833}
{Nat. Phys. {\bf 12}, 1027 (2016)}.

%%NO. 32
\bibitem{PhysRevB.92.205310}
Q. Xu, Z. Song, S. Nie, H. Weng, Z. Fang, and X. Dai,
{\emph{Two-dimensional oxide topological insulator with iron-pnictide superconductor LiFeAs structure}},
\href{https://link.aps.org/doi/10.1103/PhysRevB.92.205310}
{Phys. Rev. B {\bf 92}, 205310 (2015)}.

%%NO. 33
\bibitem{RN35}
A.Q. Wang, X.-G. Ye, D.-P. Yu, and Z.-M. Liao,
{\emph{Topological Semimetal Nanostructures: From Properties to Topotronics}},
\href{https://doi.org/10.1021/acsnano.9b07990}
{ACS Nano {\bf 14}, 3755 (2020)}.

%%No. 34    %Ref. 14  205-217
\bibitem{zhang2009topological}
H. Zhang, C.-X. Liu, X.-L. Qi, X. Dai, Z. Fang, and S.-C. Zhang,
{\emph{Topological Insulators in $\mathrm{Bi}_2\mathrm{Se}_3$, $\mathrm{Bi}_2\mathrm{Te}_3$ and $\mathrm{Sb}_2\mathrm{Te}_3$ with a Single Dirac Cone on the Surface}},
\href{https://doi.org/10.1038/nphys1270}
{{Nat. Phys.} {\bf{5}}, 438 (2009)}.

%%No. 35    %Ref. 17  252-262
\bibitem{hsieh2009tunable}
D. Hsieh, Y. Xia, D. Qian, L. Wray, J. Dil, F. Meier, J. Osterwalder, L. Patthey, J. Checkelsky, N. P. Ong \emph{et al.},
{\emph{A Tunable Topological Insulator in the Spin Helical Dirac Transport Regime}},
\href{https://doi.org/10.1038/nature08234}
{{Nature} {\bf{460}}, 1101 (2009)}.

%%No. 36    %Ref. 31  462-475
\bibitem{PhysRevLett.61.2015}
F. D. M. Haldane,
{\emph{Model for a Quantum Hall Effect Without Landau Levels: Condensed-Matter Realization of the "Parity Anomaly"}},
\href{https://link.aps.org/doi/10.1103/PhysRevLett.61.2015}
{{Phys. Rev. Lett.} {\bf{61}}, 2015 (1988)}.

%%No. 37    %Ref. 32  478-491
\bibitem{PhysRevLett.90.206601}
M. Onoda and N. Nagaosa,
{\emph{Quantized Anomalous Hall Effect in Two-Dimensional Ferromagnets: Quantum Hall Effect in Metals}},
\href{https://link.aps.org/doi/10.1103/PhysRevLett.90.206601}
{{Phys. Rev. Lett.} {\bf{90}}, 206601 (2003)}.

%%No. 38    %Ref. 33  494-507
\bibitem{PhysRevLett.101.146802}
C.-X. Liu, X.-L. Qi, X. Dai, Z. Fang, and S.-C. Zhang,
{\emph{Quantum Anomalous Hall Effect in ${\mathrm{Hg}}_{1\ensuremath{-}y}{\mathrm{Mn}}_{y}\mathrm{Te}$ Quantum Wells}},
\href{https://link.aps.org/doi/10.1103/PhysRevLett.101.146802}
{{Phys. Rev. Lett.} {\bf{101}}, 146802 (2008)}.

%%No. 39    %Ref. 35  524-535
\bibitem{AdvPhys.64.227}
H. Weng, R. Yu, X. Hu, X. Dai, and Z. Fang,
{\emph{Quantum Anomalous Hall Effect and Related Topological Electronic States}},
\href{https://doi.org/10.1080/00018732.2015.1068524}
{{Adv. Phys.} {\bf{64}}, 227 (2015)}.

%%No. 40    %Ref. 36  538-548
\bibitem{science.1187485}
R. Yu, W. Zhang, H.-J. Zhang, S.-C. Zhang, X. Dai, and Z. Fang,
{\emph{Quantized Anomalous Hall Effect in Magnetic Topological Insulators}},
\href{https://www.science.org/doi/abs/10.1126/science.1187485}
{{Science} {\bf{329}}, 61 (2010)}.

%%No. 41    %Ref. 4  50-67
\bibitem{xia2009observation}
Y. Xia, D. Qian, D. Hsieh, L. Wray, A. Pal, H. Lin, A. Bansil, D. Grauer, Y. S. Hor, R. J. Cava, and M. Z. Hasan,
{\emph{Observation of a Large-Gap Topological-Insulator Class with a Single Dirac Cone on the Surface}},
\href{https://doi.org/10.1038/nphys1274}
{{Nat. Phys.} {\bf{5}}, 398 (2009)}.

%%No. 42    %Ref. 37  551-561
\bibitem{science.1234414}
C.-Z. Chang, J. Zhang, X. Feng, J. Shen, Z. Zhang, M. Guo, K. Li, Y. Ou, P. Wei, L.-L. Wang, Z.-Q. Ji, Y. Feng, S. Ji, X. Chen, J. Jia, X. Dai, Z. Fang, S.-C. Zhang, K. He, Y. Wang, L. Lu, X.-C. Ma, and Q.-K. Xue,
{\emph{Experimental Observation of the Quantum Anomalous Hall Effect in a Magnetic Topological Insulator}},
\href{https://www.science.org/doi/abs/10.1126/science.1234414}
{{Science} {\bf{340}}, 167 (2013)}.

%%No. 43    %Ref. 38  564-576
\bibitem{ge2020high}
J. Ge, Y. Liu, J. Li, H. Li, T. Luo, Y. Wu, Y. Xu, and J. Wang,
{\emph{High-Chern-Number and High-Temperature Quantum Hall Effect Without Landau Levels}},
\href{https://doi.org/10.1093/nsr/nwaa089}
{{Natl. Sci. Rev.} {\bf{7}}, 1280 (2020)}.



%%No. 44    %Ref. 80  1182-1195
\bibitem{PhysRevLett.106.127004}
M. Kriener, K. Segawa, Z. Ren, S. Sasaki, and Y. Ando,
{\emph{Bulk Superconducting Phase with a Full Energy Gap in the Doped Topological Insulator ${\mathrm{Cu}}_{x}{\mathrm{Bi}}_{2}{\mathrm{Se}}_{3}$}},
\href{https://link.aps.org/doi/10.1103/PhysRevLett.106.127004}
{{Phys. Rev. Lett.} {\bf{106}}, 127004 (2011)}.

%%No. 45    %Ref. 44  654-666
\bibitem{wray2010observation}
L. A. Wray, S.-Y. Xu, Y. Xia, Y. S. Hor, D. Qian, A. V. Fedorov, H. Lin, A. Bansil, R. J. Cava, and M. Z. Hasan,
{\emph{Observation of Topological Order in a Superconducting Doped Topological Insulator}},
\href{https://doi.org/10.1038/nphys1762}
{{Nat. Phys.} {\bf{6}}, 855 (2010)}.



%%No. 46    %Ref. 81  1198-1210
\bibitem{Zhang24}
J. L. Zhang, S. J. Zhang, H. M. Weng, W. Zhang, L. X. Yang, Q. Q. Liu, S. M. Feng, X. C. Wang, R. C. Yu, L. Z. Cao, L. Wang, W. G. Yang, H. Z. Liu, W. Y. Zhao, S. C. Zhang, X. Dai, Z. Fang, and C. Q. Jin,
{\emph{Pressure-Induced Superconductivity in Topological Parent Compound $\mathrm{Bi}_2\mathrm{Te}_3$}},
\href{https://www.pnas.org/content/108/1/24}
{{Proc. Natl. Acad. Sci.} {\bf{108}}, 24 (2011)}.

%%No. 47    %Ref. 82  1213-1223
\bibitem{science.1216466}
M.-X. Wang, C. Liu, J.-P. Xu, F. Yang, L. Miao, M.-Y. Yao, C. L. Gao, C. Shen, X. Ma, X. Chen, Z.-A. Xu, Y. Liu, S.-C. Zhang, D. Qian, J.-F. Jia, and Q.-K. Xue,
{\emph{The Coexistence of Superconductivity and Topological Order in the $\mathrm{Bi}_2\mathrm{Se}_3$ Thin Films}},
\href{https://www.science.org/doi/abs/10.1126/science.1216466}
{{Science} {\bf{336}}, 52 (2012)}.


%%No. 48    %Ref. 18  265-275
\bibitem{science.1173034}
Y. L. Chen, J. G. Analytis, J.-H. Chu, Z. K. Liu, S.-K. Mo, X. L. Qi, H. J. Zhang, D. H. Lu, X. Dai, Z. Fang, S. C. Zhang, I. R. Fisher, Z. Hussain, and Z.-X. Shen,
{\emph{Experimental Realization of a Three-Dimensional Topological Insulator, $\mathrm{Bi}_2\mathrm{Te}_3$}},
\href{https://www.science.org/doi/abs/10.1126/science.1173034}
{{Science} {\bf{325}}, 178 (2009)}.

%%No. 49    %Ref. 19  278-291
\bibitem{PhysRevLett.103.146401}
D. Hsieh, Y. Xia, D. Qian, L. Wray, F. Meier, J. H. Dil, J. Osterwalder, L. Patthey, A. V. Fedorov, H. Lin, A. Bansil, D. Grauer, Y. S. Hor, R. J. Cava, and M. Z. Hasan,
{\emph{Observation of Time-Reversal-Protected Single-Dirac-Cone Topological-Insulator States in ${\mathrm{Bi}}_{2}{\mathrm{Te}}_{3}$ and ${\mathrm{Sb}}_{2}{\mathrm{Te}}_{3}$}},
\href{https://link.aps.org/doi/10.1103/PhysRevLett.103.146401}
{{Phys. Rev. Lett.} {\bf{103}}, 146401 (2009)}.

%%No. 50    %Ref. 20  294-306
\bibitem{arakane2012tunable}
T. Arakane, T. Sato, S. Souma, K. Kosaka, K. Nakayama, M. Komatsu, T. Takahashi, Z. Ren, K. Segawa, and Y. Ando,
{\emph{Tunable Dirac Cone in the Topological Insulator $\mathrm{Bi}_{2\ensuremath{-}x}\mathrm{Sb}_{x}\mathrm{Te}_{3\ensuremath{-}y}\mathrm{Se}_{y}$}},
\href{https://doi.org/10.1038/ncomms1639}
{{Nat. Commun.} {\bf{3}}, 636 (2012)}.



%%No. 51    %Ref. 23  337-347
\bibitem{science.1148047}
M. K{\"o}nig, S. Wiedmann, C. Br{\"u}ne, A. Roth, H. Buhmann, L. W. Molenkamp, X.-L. Qi, and S.-C. Zhang,
{\emph{Quantum Spin Hall Insulator State in $\mathrm{HgTe}$ Quantum Wells}},
\href{https://www.science.org/doi/abs/10.1126/science.1148047}
{{Science} {\bf{318}}, 766 (2007)}.

%%No. 52    %Ref. 10  145-158
\bibitem{PhysRevB.76.045302}
L. Fu and C. L. Kane,
{\emph{Topological Insulators with Inversion Symmetry}},
\href{https://link.aps.org/doi/10.1103/PhysRevB.76.045302}
{{Phys. Rev. B} {\bf{76}}, 045302 (2007)}.

%%No. 53    %Ref. 12  177-189
\bibitem{hsieh2008topological}
D. Hsieh, D. Qian, L. Wray, Y. Xia, Y. S. Hor, R. J. Cava, and M. Z. Hasan,
{\emph{A Topological Dirac Insulator in a Quantum Spin Hall Phase}},
\href{https://doi.org/10.1038/nature06843}
{{Nature} {\bf{452}}, 970 (2008)}.





%%No. 54    %Ref. 50  745-755
\bibitem{science.1245085}
Z. K. Liu, B. Zhou, Y. Zhang, Z. J. Wang, H. M. Weng, D. Prabhakaran, S.-K. Mo, Z. X. Shen, Z. Fang, X. Dai, Z. Hussain, and Y. L. Chen,
{\emph{Discovery of a Three-Dimensional Topological Dirac Semimetal, $\mathrm{Na}_3\mathrm{Bi}$}},
\href{https://www.science.org/doi/abs/10.1126/science.1245085}
{{Science} {\bf{343}}, 864 (2014)}.

%%No. 55    %Ref. 51  758-768
\bibitem{science.1256742}
S.-Y. Xu, C. Liu, S. K. Kushwaha, R. Sankar, J. W. Krizan, I. Belopolski, M. Neupane, G. Bian, N. Alidoust, T.-R. Chang, H.-T. Jeng, C.-Y. Huang, W.-F. Tsai, H. Lin, P. P. Shibayev, F.-C. Chou, R. J. Cava, and M. Z. Hasan,
{\emph{Observation of Fermi Arc Surface States in a Topological Metal}},
\href{https://www.science.org/doi/abs/10.1126/science.1256742}
{{Science} {\bf{347}}, 294 (2015)}.

%%No. 56    %Ref. 52  771-781
\bibitem{yi2014evidence}
H. Yi, Z. Wang, C. Chen, Y. Shi, Y. Feng, A. Liang, Z. Xie, S. He, J. He, Y. Peng \emph{et al.},
{\emph{Evidence of Topological Surface State in Three-Dimensional Dirac Semimetal $\mathrm{Cd}_3\mathrm{As}_2$}},
\href{https://doi.org/10.1038/srep06106}
{{Sci. Rep.} {\bf{4}}, 1 (2014)}.

%%No. 57    %Ref. 53  784-796
\bibitem{jeon2014landau}
S. Jeon, B. B. Zhou, A. Gyenis, B. E. Feldman, I. Kimchi, A. C. Potter, Q. D. Gibson, R. J. Cava, A. Vishwanath, and A. Yazdani,
{\emph{Landau Quantization and Quasiparticle Interference in the Three-Dimensional Dirac Semimetal $\mathrm{Cd}_3\mathrm{As}_2$}},
\href{https://doi.org/10.1038/nmat4023}
{{Nat. Mater.} {\bf{13}}, 851 (2014)}.

%%No. 58    %Ref. 54  799-812
\bibitem{PhysRevLett.113.246402}
L. P. He, X. C. Hong, J. K. Dong, J. Pan, Z. Zhang, J. Zhang, and S. Y. Li,
{\emph{Quantum Transport Evidence for the Three-Dimensional Dirac Semimetal Phase in ${\mathrm{Cd}}_{3}{\mathrm{As}}_{2}$}},
\href{https://link.aps.org/doi/10.1103/PhysRevLett.113.246402}
{{Phys. Rev. Lett.} {\bf{113}}, 246402 (2014)}.

%%No. 59    %Ref. 57  847-860
\bibitem{PhysRevX.5.011029}
H. Weng, C. Fang, Z. Fang, B. A. Bernevig, and X. Dai,
{\emph{Weyl Semimetal Phase in Noncentrosymmetric Transition-Metal Monophosphides}},
\href{https://link.aps.org/doi/10.1103/PhysRevX.5.011029}
{{Phys. Rev. X} {\bf{5}}, 011029 (2015)}.

%%No. 60    %Ref. 58  863-876
\bibitem{PhysRevX.5.031023}
X. Huang, L. Zhao, Y. Long, P. Wang, D. Chen, Z. Yang, H. Liang, M. Xue, H. Weng, Z. Fang, X. Dai, and G. Chen,
{\emph{Observation of the Chiral-Anomaly-Induced Negative Magnetoresistance in 3D Weyl Semimetal $\mathrm{Ta}\mathrm{As}$}},
\href{https://link.aps.org/doi/10.1103/PhysRevX.5.031023}
{{Phys. Rev. X} {\bf{5}}, 031023 (2015)}.



%%No. 61    %Ref. 47  700-710
\bibitem{schewe1989crystal}
I. Schewe, P. B{\"o}ttcher, and H. v. Schnering,
{\emph{The Crystal Structure of $\mathrm{Tl}_5\mathrm{Te}_3$ and its Relationship to the $\mathrm{Cr}_5\mathrm{B}_3$ Type}},
\href{https://doi.org/10.1524/zkri.1989.188.3-4.287}
{{Zeitschrift f{\"u}r Kristallographie} {\bf{188}}, 287 (1989)}.



%%No. 62    %Ref. 61  909-919
\bibitem{bradtmoller1994crystal}
S. Bradtm{\"o}ller and P. B{\"o}ttcher,
{\emph{Crystal Structure of Molybdenum Tetrathallium Tritelluride, $\mathrm{Mo}\mathrm{Tl}_{4}\mathrm{Te}_{3}$}},
\href{https://doi.org/10.1524/zkri.1994.209.1.75}
{{Zeitschrift f{\"u}r Kristallographie-Crystalline Materials} {\bf{209}}, 75 (1994)}.

%%No. 63    %Ref. 62  922-932
\bibitem{imamalieva2008new}
S. Imamalieva, F. Sadygov, and M. Babanly,
{\emph{New Thallium Neodymium Tellurides}},
\href{https://doi.org/10.1134/S0020168508090070}
{{Inorg. Mater.} {\bf{44}}, 935 (2008)}.

%%No. 64    %Ref. 34  510-521
\bibitem{PhysicsandChemistryofSolidState.21.492}
 ImamaliyevaS.,
{\emph{$\mathrm{Tl}_4\mathrm{Gd}\mathrm{Te}_3$ and $\mathrm{Tl}_4\mathrm{Dy}\mathrm{Te}_3$ - Novel Structural $\mathrm{Tl}_5\mathrm{Te}_3$ Analogues}},
\href{https://comp-sc.pu.if.ua/index.php/pcss/article/view/4293}
{{Physics and Chemistry of Solid State} {\bf{21}}, 492 (2020)}.

%%No. 65
\bibitem{RN34}
S. Z. Imamaliyeva, D. M. Babanly, D. B. Tagiev, and M. B. Babanly,
{\emph{Physicochemical Aspects of Development of Multicomponent Chalcogenide Phases Having the Tl5Te3 Structure: A Review}},
\href{https://doi.org/10.1134/S0036023618130041}
{Russian Journal of Inorganic Chemistry {\bf 63}, 1704 (2018)}.

%%No. 66    %Ref. 76  1123-1133
\bibitem{bradtmoller1993darstellung}
S. Bradtm{\"o}ller and P. B{\"o}ttcher,
{\emph{Darstellung Und Kristallstruktur Von $\mathrm{Sn}\mathrm{Tl}_4\mathrm{Te}_3$ Und $\mathrm{Pb}\mathrm{Tl}_{4}\mathrm{Te}_{3}$}},
\href{https://doi.org/10.1002/zaac.19936190702}
{{Zeitschrift f{\"u}r anorganische und allgemeine Chemie} {\bf{619}}, 1155 (1993)}.




%%No. 67    %Ref. 46  684-697
\bibitem{PhysRevLett.112.017002}
K. E. Arpino, D. C. Wallace, Y. F. Nie, T. Birol, P. D. C. King, S. Chatterjee, M. Uchida, S. M. Koohpayeh, J.-J. Wen, K. Page, C. J. Fennie, K. M. Shen, and T. M. McQueen,
{\emph{Evidence for Topologically Protected Surface States and a Superconducting Phase in $[{\mathrm{Tl}}_{4}]({\mathrm{Tl}}_{1\ensuremath{-}x}{\mathrm{Sn}}_{x}){\mathrm{Te}}_{3}$Using Photoemission, Specific Heat, and Magnetization Measurements, and Density Functional Theory}},
\href{https://link.aps.org/doi/10.1103/PhysRevLett.112.017002}
{{Phys. Rev. Lett.} {\bf{112}}, 017002 (2014)}.

%%No. 68    %Ref. 83  1226-1236
\bibitem{APLMaterials.3.041507}
K. E. Arpino, B. D. Wasser, and T. M. McQueen,
{\emph{Superconducting Dome and Crossover to an Insulating State in $\mathrm{[}\mathrm{Tl}_4\mathrm{]}\mathrm{Tl}_{1\ensuremath{-}x}\mathrm{Sn}_x\mathrm{Te}_3$}},
\href{https://doi.org/10.1063/1.4913392}
{{APL Materials} {\bf{3}}, 041507 (2015)}.








%%%No. 59    %Ref. 63  935-945
%\bibitem{imamaliyeva2018physicochemical}
%S. Imamaliyeva, D. Babanly, D. Tagiev, and M. Babanly,
%{\emph{Physicochemical Aspects of Development of Multicomponent Chalcogenide Phases Having the $\mathrm{Tl}_5\mathrm{Te}_3$ Structure: A Review}},
%\href{https://doi.org/10.1134/S0036023618130041}
%{{Russ. J. Inorg. Chem.} {\bf{63}}, 1704 (2018)}.

%%Ref. 30    390-401
\bibitem{SETYAWAN2010299}
W. Setyawan and S. Curtarolo,
{High-throughput electronic band structure calculations: Challenges and tools},
\href{https://www.sciencedirect.com/science/article/pii/S0927025610002697}
{Comput. Mater. Sci. {\bf 49}, 299 (2010)}.



%%No. 69    %Ref. 65  964-975
\bibitem{KRESSE199615}
G. Kresse and J. Furthm{\"u}ller,
{\emph{Efficiency of Ab-Initio Total Energy Calculations for Metals and Semiconductors Using a Plane-Wave Basis Set}},
\href{https://www.sciencedirect.com/science/article/pii/0927025696000080}
{{Comput. Mater. Sci.} {\bf{6}}, 15 (1996)}.

%%No. 70    %Ref. 64  948-961
\bibitem{PhysRevB.54.11169}
G. Kresse and J. Furthm\"uller,
{\emph{Efficient Iterative Schemes for Ab Initio Total-Energy Calculations Using a Plane-Wave Basis Set}},
\href{https://link.aps.org/doi/10.1103/PhysRevB.54.11169}
{{Phys. Rev. B} {\bf{54}}, 11169 (1996)}.



%%No. 71    %Ref. 66  978-991
\bibitem{PhysRevB.50.17953}
P. E. Bl\"ochl,
{\emph{Projector Augmented-Wave Method}},
\href{https://link.aps.org/doi/10.1103/PhysRevB.50.17953}
{{Phys. Rev. B} {\bf{50}}, 17953 (1994)}.


%% 72 LDA
\bibitem{PhysRevB.23.5048}
J. P. Perdew and A. Zunger,
{Self-interaction correction to density-functional approximations for many-electron systems},
\href{https://link.aps.org/doi/10.1103/PhysRevB.23.5048}
{Phys. Rev. B {\bf 23}, 5048 (1981)}.



%%%No. 72  PBE  %Ref. 67  994-1007
%\bibitem{PhysRevLett.77.3865}
%J. P. Perdew, K. Burke, and M. Ernzerhof,
%{\emph{Generalized Gradient Approximation Made Simple}},
%\href{https://link.aps.org/doi/10.1103/PhysRevLett.77.3865}
%{{Phys. Rev. Lett.} {\bf{77}}, 3865 (1996)}.

%%No. 73    %Ref. 71  1052-1063
\bibitem{TOGO20151}
A. Togo and I. Tanaka,
{\emph{First Principles Phonon Calculations in Materials Science}},
\href{https://www.sciencedirect.com/science/article/pii/S1359646215003127}
{{Scr. Mater.} {\bf{108}}, 1 (2015)}.

%%No. 74
\bibitem{PhysRevB.55.10355}
X. Gonze and C. Lee,
{\emph{Dynamical matrices, Born effective charges, dielectric permittivity tensors, and interatomic force constants from  density-functional perturbation theory}},
\href{https://link.aps.org/doi/10.1103/PhysRevB.55.10355}
{Phys. Rev. B {\bf 55}, 10355 (1997)}.

%%No. 75
\bibitem{PhysRevLett.102.226401}
F. Tran and P. Blaha,
{\emph{Accurate Band Gaps of Semiconductors and Insulators with a Semilocal Exchange-Correlation Potential}},
\href{https://link.aps.org/doi/10.1103/PhysRevLett.102.226401}
{Phys. Rev. Lett. {\bf 102}, 226401 (2009)}.


%%Ref. 76    261-274
\bibitem{PhysRevB.85.155109}
D. Koller, F. Tran, and P. Blaha,
{\emph{Improving the modified Becke-Johnson exchange potential}},
\href{https://link.aps.org/doi/10.1103/PhysRevB.85.155109}
{Phys. Rev. B {\bf 85}, 155109 (2012)}.

%%No. 84    %Ref. 68  1010-1020
\bibitem{heyd2003hybrid}
J. Heyd, G. E. Scuseria, and M. Ernzerhof,
{\emph{Hybrid Functionals Based on a Screened Coulomb Potential}},
\href{https://doi.org/10.1063/1.1564060}
{{J. Chem. Phys.} {\bf{118}}, 8207 (2003)}.

%%No. 85
\bibitem{PhysRev.139.A796}
L. Hedin,
{New Method for Calculating the One-Particle Green's Function with Application to the Electron-Gas Problem},
\href{https://link.aps.org/doi/10.1103/PhysRev.139.A796}
{Phys. Rev. {\bf 139}, A796 (1965)}.

%%Ref. 77    118-131
\bibitem{PhysRevB.83.195134}
D. Koller, F. Tran, and P. Blaha,
{\emph{Merits and limits of the modified Becke-Johnson exchange potential}},
\href{https://link.aps.org/doi/10.1103/PhysRevB.83.195134}
{Phys. Rev. B {\bf 83}, 195134 (2011)}.

%%Ref. 78    132-145
\bibitem{PhysRevB.82.155145}
D. J. Singh,
{\emph{Structure and optical properties of high light output halide scintillators}},
\href{https://link.aps.org/doi/10.1103/PhysRevB.82.155145}
{Phys. Rev. B {\bf 82}, 155145 (2010)}.

%%Ref. 79    146-159
\bibitem{PhysRevB.82.205212}
Y.-S. Kim, M. Marsman, G. Kresse, F. Tran, and P. Blaha,
{\emph{Towards efficient band structure and effective mass calculations for III-V direct band-gap semiconductors}},
\href{https://link.aps.org/doi/10.1103/PhysRevB.82.205212}
{Phys. Rev. B {\bf 82}, 205212 (2010)}.

%%Ref. 80    160-173
\bibitem{PhysRevB.82.205102}
D. J. Singh,
{\emph{Electronic structure calculations with the Tran-Blaha modified Becke-Johnson density functional}},
\href{https://link.aps.org/doi/10.1103/PhysRevB.82.205102}
{Phys. Rev. B {\bf 82}, 205102 (2010)}.

%%Ref. 81    174-187
\bibitem{PhysRevB.93.045304}
S. K\"ufner, L. Matthes, and F. Bechstedt,
{\emph{Quantum spin Hall effect in $\ensuremath{\alpha}\ensuremath{-}\mathrm{Sn}/\text{CdTe}(001)$ quantum-well structures}},
\href{https://link.aps.org/doi/10.1103/PhysRevB.93.045304}
{Phys. Rev. B {\bf 93}, 045304 (2016)}.

%%Ref. 82    188-201
\bibitem{PhysRevB.93.115104}
J. Lee, A. Seko, K. Shitara, K. Nakayama, and I. Tanaka,
{\emph{Prediction model of band gap for inorganic compounds by combination of density functional theory calculations and machine learning techniques}},
\href{https://link.aps.org/doi/10.1103/PhysRevB.93.115104}
{Phys. Rev. B {\bf 93}, 115104 (2016)}.

%%Ref. 83    202-215
\bibitem{PhysRevB.101.245163}
Tom\'a\ifmmode \check{s}\else \v{s}\fi{} Rauch, M. A. L. Marques, and S. Botti,
{\emph{Accurate electronic band gaps of two-dimensional materials from the local modified Becke-Johnson potential}},
\href{https://link.aps.org/doi/10.1103/PhysRevB.101.245163}
{Phys. Rev. B {\bf 101}, 245163 (2020)}.




%%No. 86    %Ref. 69  1023-1035
\bibitem{MOSTOFI20142309}
A. A. Mostofi, J. R. Yates, G. Pizzi, Y.-S. Lee, I. Souza, D. Vanderbilt, and N. Marzari,
{\emph{An Updated Version of Wannier90: A Tool for Obtaining Maximally-Localised Wannier Functions}},
\href{https://www.sciencedirect.com/science/article/pii/S001046551400157X}
{{Comput. Phys. Commun.} {\bf{185}}, 2309 (2014)}.

%%No. 87    %Ref. 70  1038-1049
\bibitem{WU2018405}
Q. Wu, S. Zhang, H.-F. Song, M. Troyer, and A. A. Soluyanov,
{\emph{WannierTools: An Open-Source Software Package for Novel Topological Materials}},
\href{https://www.sciencedirect.com/science/article/pii/S0010465517303442}
{{Comput. Phys. Commun.} {\bf{224}}, 405 (2018)}.

%%No. 88
\bibitem{RN36}
J. M. Clary, A. M. Holder, and C. B. Musgrave,
{\emph{Computationally Predicted High-Throughput Free-Energy Phase Diagrams for the Discovery of Solid-State Hydrogen Storage Reactions}},
\href{https://doi.org/10.1021/acsami.0c13298}
{ACS Applied Materials $\&$ Interfaces {\bf 12}, 48553 (2020)}.

%%Ref. 89    66-77
\bibitem{IntegratedFerroelectrics.175.186}
A. P. Jaroenjittichai,
{\emph{Formation energies and chemical potential diagrams of II-Ge-N2 semiconductors}},
\href{https://doi.org/10.1080/10584587.2016.1204202}
{Integrated Ferroelectrics {\bf 175}, 186 (2016)}.

%%No. 90    %Ref. 60  894-906
\bibitem{ma13194303}
D. Mutter, D. F. Urban, and C. Els{\"a}sser,
{\emph{Determination of Formation Energies and Phase Diagrams of Transition Metal Oxides with DFT+U}},
\href{https://www.mdpi.com/1996-1944/13/19/4303}
{{Materials} {\bf{13}}, 4303 (2020)}.

%%No. 91
\bibitem{JournaloftheAmericanCeramicSociety.72.2104}
H. Yokokawa, T. Kawada, and M. Dokiya,
{\emph{Construction of Chemical Potential Diagrams for Metal-Metal-Nonmetal Systems: Applications to the Decomposition of Double Oxides}},
\href{https://ceramics.onlinelibrary.wiley.com/doi/abs/10.1111/j.1151-2916.1989.tb06039.x}
{Journal of the American Ceramic Society {\bf 72}, 2104 (1989)}.

%%No. 92
\bibitem{RN21}
H. Yokokawa,
{\emph{Generalized chemical potential diagram and its applications to chemical reactions at interfaces between dissimilar materials}},
\href{https://doi.org/10.1361/105497199770335794}
{Journal of Phase Equilibria {\bf 20}, 258 (1999)}.

%%No.93
\bibitem{RN22}
K. T. Jacob, T. H. Okabe, T. Uda, and Y. Waseda,
{\emph{System Cu-Rh-O: Phase diagram and thermodynamic properties of ternary oxides CuRhO2 and CuRh2O4}},
\href{https://doi.org/10.1007/BF02745598}
{Bulletin of Materials Science {\bf 22}, 741 (1999)}.


%%No.94
\bibitem{SM}
See Supplemental Material at
\href{https://www.baidu.com}
{URL will be inserted later}
for detials.

%%Ref. 26    341-351
\bibitem{JChemPhys.154.234706}
A. Sharan and S. Lany,
{Computational discovery of stable and metastable ternary oxynitrides},
\href{https://doi.org/10.1063/5.0050356}
{J. Chem. Phys. {\bf 154}, 234706 (2021)}.

%%Ref. 27    352-365
\bibitem{PhysRevLett.123.097001}
Y. Sun, J. Lv, Y. Xie, H. Liu, and Y. Ma,
{Route to a Superconducting Phase above Room Temperature in Electron-Doped Hydride Compounds under High Pressure},
\href{https://link.aps.org/doi/10.1103/PhysRevLett.123.097001}
{Phys. Rev. Lett. {\bf 123}, 097001 (2019)}.

%%%Ref. 28    366-376
%\bibitem{sun2017thermodynamic}
%W. Sun, A. Holder, B. Orva{\ n}anos, E. Arca, A. Zakutayev, S. Lany, and G. Ceder,
%{Thermodynamic routes to novel metastable nitrogen-rich nitrides},
%\href{https://doi.org/10.1021/acs.chemmater.7b02399}
%{Chem. Mater. {\bf 29}, 6936 (2017)}.
%
%%%Ref. 29    377-389
%\bibitem{ye2022}
%W. Ye, X. Lei, M. Aykol, and J. H. Montoya,
%{Novel inorganic crystal structures predicted using autonomous simulation agents},
%\href{https://doi.org/10.1038/s41597-022-01438-8}
%{Scientific Data {\bf 9}, 302 (2022)}.


%%No.95
\bibitem{GUSEINOV1969807}
G. Guseinov, G. Abdullayev, E. Kerimova, R. Gamidov, and G. Guseinov,
{Structural and physical properties of three-component CdInS2 (Se2,Te2) compounds},
\href{https://www.sciencedirect.com/science/article/pii/0025540869900038}
{Materials Research Bulletin {\bf 4}, 807 (1969)}.

%%No. 96    %Ref. 59  879-891
\bibitem{narang2021topology}
P. Narang, C. A. C. Garcia, and C. Felser,
{\emph{The Topology of Electronic Band Structures}},
\href{https://doi.org/10.1038/s41563-020-00820-4}
{{Nat. Mater.} {\bf{20}}, 293 (2021)}.

%%No. 97
\bibitem{PhysRevLett.98.106803}
L. Fu, C. L. Kane, and E. J. Mele,
{Topological Insulators in Three Dimensions},
\href{https://link.aps.org/doi/10.1103/PhysRevLett.98.106803}
{Phys. Rev. Lett. {\bf 98}, 106803 (2007)}.

%%No. 98    %Ref. 78  1151-1164
\bibitem{PhysRevB.84.075119}
R. Yu, X. L. Qi, A. Bernevig, Z. Fang, and X. Dai,
{\emph{Equivalent Expression of ${\mathbb{Z}}_{2}$ Topological Invariant for Band Insulators Using the Non-Abelian Berry Connection}},
\href{https://link.aps.org/doi/10.1103/PhysRevB.84.075119}
{{Phys. Rev. B} {\bf{84}}, 075119 (2011)}.

%%No. 99    %Ref. 1  5-17
\bibitem{JPhysFMetPhys.14.1205}
M. P. L. Sancho, J. M. L. Sancho, and J. Rubio,
{\emph{Quick Iterative Scheme for the Calculation of Transfer Matrices: Application to Mo (100)}},
\href{https://doi.org/10.1088/0305-4608/14/5/016}
{{J. Phys. F: Met. Phys.} {\bf{14}}, 1205 (1984)}.

%%No. 100    %Ref. 2  20-32
\bibitem{JPhysFMetPhys.15.851}
M. P. L. Sancho, J. M. L. Sancho, J. M. L. Sancho, and J. Rubio,
{\emph{Highly Convergent Schemes for the Calculation of Bulk and Surface Green Functions}},
\href{https://doi.org/10.1088/0305-4608/15/4/009}
{{J. Phys. F: Met. Phys.} {\bf{15}}, 851 (1985)}.

%%No.101
\bibitem{Nie10596}
S. Nie, G. Xu, F. B. Prinz, and S.-c. Zhang,
{\emph{Topological semimetal in honeycomb lattice LnSI}},
\href{https://www.pnas.org/content/114/40/10596}
{Proc. Natl. Acad. Sci. {\bf 114}, 10596 (2017)}.


%%No. 78    %Ref. 77  1136-1148
%\bibitem{Nie10596}
%S. Nie, G. Xu, F. B. Prinz, and S.-c. Zhang,
%{\emph{Topological Semimetal in Honeycomb Lattice $\mathrm{LnSI}$}},
%\href{https://www.pnas.org/content/114/40/10596}
%{{Proc. Natl. Acad. Sci.} {\bf{114}}, 10596 (2017)}.

%%No. 79    %Ref. 3  35-47
%\bibitem{RussianChemicalReviews.77.1}
%A. V. Shevelkov,
%{\emph{Chemical Aspects of the Design of Thermoelectric Materials}},
%\href{https://doi.org/10.1070/rc2008v077n01abeh003746}
%{{Russian Chemical Reviews} {\bf{77}}, 1 (2008)}.

%%No. 80    %Ref. 24  350-363
%\bibitem{PhysRevLett.95.146802}
%C. L. Kane and E. J. Mele,
%{\emph{${Z}_{2}$ Topological Order and the Quantum Spin Hall Effect}},
%\href{https://link.aps.org/doi/10.1103/PhysRevLett.95.146802}
%{{Phys. Rev. Lett.} {\bf{95}}, 146802 (2005)}.

%%No. 81    %Ref. 25  366-379
%\bibitem{PhysRevLett.95.226801}
%C. L. Kane and E. J. Mele,
%{\emph{Quantum Spin Hall Effect in Graphene}},
%\href{https://link.aps.org/doi/10.1103/PhysRevLett.95.226801}
%{{Phys. Rev. Lett.} {\bf{95}}, 226801 (2005)}.

%%No. 82    %Ref. 26  382-395
%\bibitem{PhysRevB.79.195321}
%R. Roy,
%{\emph{${Z}_{2}$ Classification of Quantum Spin Hall Systems: An Approach Using Time-Reversal Invariance}},
%\href{https://link.aps.org/doi/10.1103/PhysRevB.79.195321}
%{{Phys. Rev. B} {\bf{79}}, 195321 (2009)}.

%%No. 83    %Ref. 27  398-411
%\bibitem{PhysRevB.79.195322}
%R. Roy,
%{\emph{Topological Phases and the Quantum Spin Hall Effect in Three Dimensions}},
%\href{https://link.aps.org/doi/10.1103/PhysRevB.79.195322}
%{{Phys. Rev. B} {\bf{79}}, 195322 (2009)}.

%%No. 84    %Ref. 45  669-681
%\bibitem{NORDELL199651}
%K. J. Nordell and G. J. Miller,
%{\emph{Electronic Structure, Superconductivity, and Substitution Patterns in $\mathrm{Tl}_5\mathrm{Te}_3$}},
%\href{https://www.sciencedirect.com/science/article/pii/0925838896023031}
%{{J. Alloys Compd.} {\bf{241}}, 51 (1996)}.

%%No. 85    %Ref. 72  1066-1079
%\bibitem{PhysRevLett.112.096804}
%H. Zhang, J. Wang, G. Xu, Y. Xu, and S.-C. Zhang,
%{\emph{Topological States in Ferromagnetic $\mathrm{CdO}/\mathrm{EuO}$ Superlattices and Quantum Wells}},
%\href{https://link.aps.org/doi/10.1103/PhysRevLett.112.096804}
%{{Phys. Rev. Lett.} {\bf{112}}, 096804 (2014)}.

%%No. 86    %Ref. 73  1082-1091
%\bibitem{1386971}
%K. Butler, J. Saxena, A. Jain, T. Fryars, J. Lewis, and G. Hetherington,
%{\emph{Minimizing Power Consumption in Scan Testing: Pattern Generation and DFT Techniques}},
%\href{http://dx.doi.org/10.1109/TEST.2004.1386971}
%{{2004 International Conferce on Test} {\bf{}}, 355 (2004)}.

%%No. 87    %Ref. 74  1094-1107
%\bibitem{PhysRev.139.A796}
%L. Hedin,
%{\emph{New Method for Calculating the One-Particle Green's Function with Application to the Electron-Gas Problem}},
%\href{https://link.aps.org/doi/10.1103/PhysRev.139.A796}
%{{Phys. Rev.} {\bf{139}}, A796 (1965)}.

%%No. 88    %Ref. 75  1110-1120
%\bibitem{karlicky2013band}
%F. Karlicky and M. Otyepka,
%{\emph{Band Gaps and Optical Spectra of Chlorographene, Fluorographene and Graphane from G0W0, GW0 and GW Calculations on Top of PBE and HSE06 Orbitals}},
%\href{https://doi.org/10.1021/ct400476r}
%{{J. Chem. Theory Comput.} {\bf{9}}, 4155 (2013)}.

%%%
%%% Uncomment the following lines for Supplemental Material
%%%



\end{thebibliography}
\end{document}